\documentclass[twocolumn,showpacs,preprintnumbers,prl,fleqn]{revtex4}
\usepackage{epsfig,graphicx,amsfonts,amsbsy}
\input{psfig.sty}
\begin{document}

\newcommand{\beqa}{\begin{eqnarray}}
\newcommand{\eeqa}{\end{eqnarray}}
\newcommand{\beq}{\begin{equation}}
\newcommand{\eeq}{\end{equation}}

\newcommand{\tr}{{\rm tr}}
\newcommand{\g}{{\rm\Large g}}

\title{Spin Transfer Without Spin Conservation}

\author{Alvaro S. N\'u\~nez and A.H. MacDonald}
\affiliation{Physics Department, University of Texas at Austin,
Austin TX 78712}

\date{\today}

\begin{abstract}

We propose a general theory of the spin-transfer effects that
occur when current flows through inhomogeneous magnetic systems.
Our theory does not rest on an appeal to conservation of total
spin, can assess whether or not
current-induced magnetization precession and switching
in a particular geometry will occur coherently, and
can estimate the efficacy of spin-transfer when spin-orbit
interactions are
present.  We illustrate our theory by applying it to a
toy-model two-dimensional-electron-gas ferromagnet with
Rashba spin-orbit interactions.

\end{abstract}

\pacs{75.50.Pp, 75.30.Ds, 73.43.-f}

\maketitle

\noindent
{\em Introduction:}
The transfer\cite{Slon,Berger} of magnetization from
quasiparticles to collective degrees of freedom
in transition metal ferromagnets has received attention recently
because of experimental\cite{Tsoi1,Tsoi2,Sun,SMT-exp,Chien,MSU} 
and theoretical\cite{TransferTheory} progress that has
motivated basic science interest in this many-electron
phenomenon, and because of the possibility that the effect might
prove to be a useful way to
write magnetic information.  The key theoretical ideas that
underly this effect were
proposed some time ago\cite{Slon,Berger} and rest heavily on
bookkeeping which follows the flow
of spin-angular momentum through the system.  Recent advances in
nanomagnetism have made it possible to compare these ideas with
experimental observations and explore them more
fully.  In this Letter, we propose a general theory of spin transfer that
does not rest on an appeal to conservation of total spin,
focusing instead on the change in the
exchange-field experienced by quasiparticles in the presence of
non-zero transport currents. 
Our approach can assess whether or not the current-driven
magnetization dynamics in a particular geometry will be coherent, and
can predict the efficacy of spin-transfer when spin-orbit
interactions are present.  It can be formulated within any time-dependent
mean field theory of a metallic ferromagnet but is, for
transition metals, most appropriately placed in the 
framework of {\em ab initio} spin-density-functional
theory\cite{Gunnarsson} (SDFT) which is accurate for these systems.

Our picture of spin-transfer is summarized schematically in Fig.[1].
In SDFT, order in a metallic ferromagnet  
is characterized by excess occupation of {\em majority}-spin orbitals, at a band energy cost  
smaller than the exchange-correlation energy gain.  (Adopting the common 
terminology of magnetism, we refer to the scalar and spin exchange-correlation fields of 
SDFT below as scalar and exchange potentials.) 
In the ordered state, majority and minority spin 
quasiparticles are brought into equilibrium by an exchange field that 
is approximately proportional to the magnetization magnitude 
and points in the majority-spin direction.  The spin-orientation of the singly
occupied majority-spin orbitals is the collective-coordinate, the magnetization 
orientation, that plays the lead role in most magnetic phenomena.
The non-equilibrium current-carrying state of a ferromagnetic metal thin film
can then be described using a scattering or non-equilibrium Greens function
formulation of transport theory\cite{Landauer}.  The current 
is due to electrons in a narrow {\em transport window} with  
width $eV$ centered on the Fermi energy, and can be evaluated by solving 
the quasiparticle Schroedinger equation for electrons incident from 
the high-potential-energy
side of the film.  The spin-transfer effect occurs when the spin-polarization 
of these transport electrons
is not parallel to the magnetization, producing a  
transport induced exchange field around which the magnetization precesses.  
We expand on this picture below and illustrate its utility by applying it to a
toy-model two-dimensional ferromagnet with Rashba\cite{Rashba}
spin-orbit interactions.

\begin{figure}
\vspace{-0.5cm}
\centerline{\psfig{figure=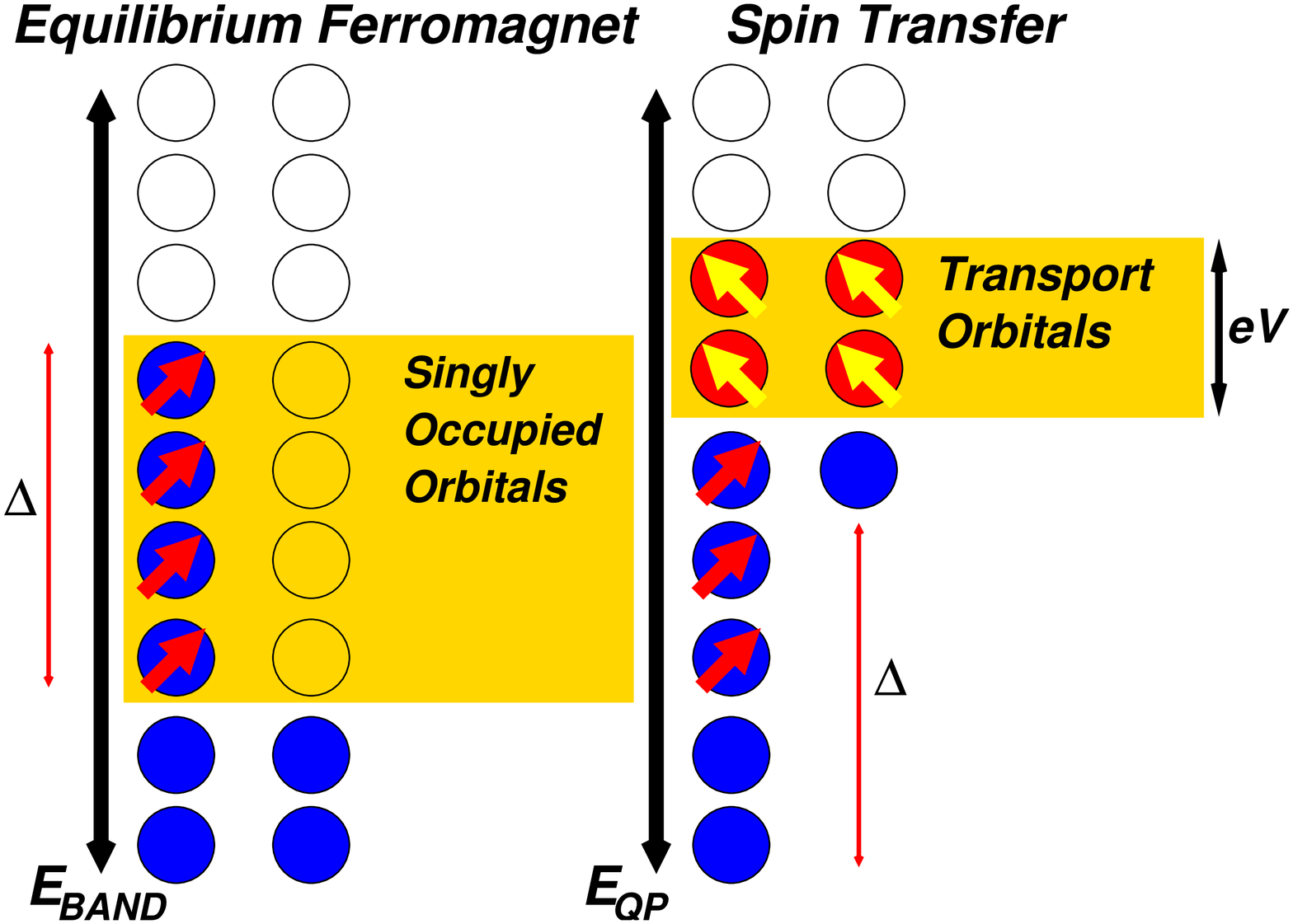,height=1.3in,width=2.4in}}
\caption{ Left panel: Ground state of a metallic ferromagnet.  The low-energy 
collective degree of freedom is the spin-orientation of singly occupied orbitals.
Right panel: Quasiparticles experience a strong exchange field $\vec{\Delta}$ that 
brings majority and minority spins into equilibrium.  Because this field is parallel
to the magnetization it does not produce a torque.  In an inhomogeneous ferromagnet, 
the spin orientation of the transport orbitals 
in a window of width $eV$ at the Fermi energy can differ from the magnetization orientation. 
The spin-transfer torque is produced by the transport-orbital
contribution to the exchange field.}
\label{fig1}
\end{figure}

\noindent
{\em Quasiparticle Spin Dynamics:}
We start by considering single-particle Hamiltonians of the form
\beq
{\cal H} = \frac{p^2}{2m} +V(\vec r) -\frac{1}{2}
\vec{\Delta}(\vec r)\cdot \vec{\tau},
\eeq
where $V(\vec r)$ and $\vec{\Delta}(\vec r)$ are arbitrary scalar
and exchange potentials
and $\vec{\tau}$ is the Pauli spin-matrix vector.  In the local-spin-density 
approximation\cite{Gunnarsson} (LSDA) of SDFT, 
${\vec \Delta}(\vec r) = \Delta_0(n(\vec r),m(\vec r)) {\hat m}(\vec r)$
where $\hat m$ is a unit vector, ${\vec m} = m\,{\hat m}$ is the total spin-density at 
$\vec r$ obtained in equilibrium by summing over all
occupied orbitals, and the magnitude of the exchange field ($\Delta_0(n,m)$) 
is the quasiparticle spin-splitting of a polarized uniform electron gas.
The spin-density contribution from a single orbital $\Psi_{\alpha}$ is
$\vec{s}_{\alpha}(\vec{r})
=  \Psi_{\alpha}^\dagger(\vec{r})\,\, \vec{\tau}\,\, \Psi_{\alpha}(\vec{r})/2$.
The time-dependent quasiparticle Schroedinger equation therefore implies that   
\beq
\frac{d s_{\alpha,j}(\vec{r})}{dt} =
\nabla_i J^i_{\alpha,j}(\vec{r}) + \frac{1}{\hbar} \left[ \vec{\Delta}\times 
\vec{s}_{\alpha}(\vec{r})\right]_j
\label{quasiparticledynamics}
\eeq
where the spin current tensor for orbital $\alpha$ is defined by,
\beq
J^i_{\alpha,j}(\vec{r})=\frac{1}{2m} {\rm Im} \left( \Psi_{\alpha}^\dagger(\vec{r})\tau_j
\nabla^i
\Psi_{\alpha}(\vec{r})\right).
\eeq
This equation exhibits the separate contributions to individual quasiparticle
spin dynamics from convective spin flow, the
source of the conservative term, and precession around the exchange field
$\vec{\Delta}$.  Both sides of Eq.[~\ref{quasiparticledynamics}] vanish when
the quasiparticle spinor solves a time-independent Schroedinger equation.

\noindent
{\em Collective Magnetization Dynamics:} 
The time-dependence of the total magnetization is obtained by summing
Eq.[~\ref{quasiparticledynamics}] over all occupied orbitals. 
\beq
\frac{d m_j(\vec{r})}{dt} =
\sum_{\alpha} \nabla_i  J^i_{\alpha,j}(\vec{r}) + \frac{1}{\hbar} \left[ \vec{\Delta}\times
\vec{m}(\vec{r})\right]_j
\label{collectivedynamics1}
\eeq
where $J^i_{\alpha,j}$ is the contribution to the spin-current from orbital $\alpha$.  
The main point we wish to make here is that (in the LSDA) $\vec{\Delta}$ is proportional to
$\vec{m}$ at each point in space-time so that the second term on the right vanishes.
The collective magnetization dynamics\cite{Vignale} is driven not by the large effective fields seen by the 
quasiparticles, but by external and demagnetization fields and spin-orbit coupling effects 
that have been neglected to this point in the discussion, and by the divergence of the 
collective spin-current\cite{KoenigSpinCurrent} in the first term.  A complete description of 
magnetization dynamics would require that the neglected terms be included, and that 
damping due to magnetophonon and other couplings be recognized.  In practice, thin film 
magnetization dynamics can usually be successfully described using a partially phenomenological 
{\em micromagnetic} theory approach\cite{Micromagnetics} in which the long-wavelength limit of the 
microscopic physics is represented by a small number of material parameters that 
specify magnetic anisotropy, stiffness, and 
damping.  We adopt that pragmatic approach here, replacing the microscopic 
Eq.[~\ref{collectivedynamics1}] by the phenomenological Landau-Liftshitz equation 
\beq
\frac{\partial \hat{m}}{\partial t}=
\hat{m}\times {\vec \Delta}^{C}+
\alpha\, \hat{m}\times\frac{\partial \hat{m}}{\partial t},
\label{collectivedynamics2} 
\eeq
where $\alpha$ is the damping parameter, 
\beq
{\vec \Delta}^{C}(\vec{r}) \equiv \frac{\delta E_{MM}[\hat{m}]}{\delta \hat{m}(\vec{r})}
\eeq
is the effective field that drives the long-wavelength collective dynamics
of an electrically isolated sample, and $E_{MM}[\hat m]$ is the micromagnetic energy functional.

\noindent
{\em Spin-Transfer:}  When current flows through a ferromagnet, the transport orbitals are few 
in number and make a negligibly small contribution to the magnitude of the magnetization.  In an inhomogeneous 
magnetic system, however, they can make an important contribution to the exchange field 
$\vec{\Delta}$ as we now explain.  Because $\vec \Delta$ is much larger than $\vec \Delta^{C}$, slow collective
dynamics can be ignored in the transport theory.  Our approach to spin-transfer is 
based on a scattering theory formulation\cite{Landauer} in which properties of interest can be expressed in 
terms of scattering solutions of the time-independent Schroedinger equation defined by the 
instantaneous value of $\vec \Delta$.  Transport electrons will in general make a contribution to the spin-density
that is small but perpendicular to the magnetization\cite{caveatonlongitudinal}.  We define this
transport contribution to the spin-density as ${\vec m}^{\rm tr}$.  Because it is perpendicular to the  
magnetization, its contribution to the exchange-field experienced by all quasiparticles 
\beq 
{\vec \Delta}^{\rm tr} = \Delta_0(n,m) \; \frac{{\vec m}^{\rm tr}}{m} 
\label{deltatr}
\eeq
produces a spin-torque that can be comparable to that produced by 
$\vec{\Delta}^{C}$. 
It follows that the influence of a transport current on magnetization dynamics is captured by
replacing ${\vec \Delta}^{C}$ in Eq.[~\ref{collectivedynamics2}] by ${\vec \Delta}^{C} + {\vec \Delta}^{\tr}$.
This proposal is the central idea of our paper.

Our proposal can be related to the present approach in which spin-transfer is computed from
spin current fluxes.  In the absence of spin-orbit coupling, summing over all transport orbitals 
and applying Eq.[~\ref{quasiparticledynamics}] implies a relationship between the transport
magnetization and the transport spin currents:
\beq
\left[ \vec{\Delta}(\vec{r}) \times 
\vec{m}^{\rm tr}(\vec{r})\right]_j = - \hbar \nabla_i J^{\tr,i}_j(\vec{r})   
\label{transportdelta}
\eeq
where $J^{{\rm tr},i}_j$ is the spin-current tensor summed over all transport orbitals.
Note that the net spin current flux through any small volume is always perpendicular to the 
magnetization. It follows from Eq.[~\ref{transportdelta}] that
\beq 
{\vec \Delta}^{\rm tr}(\vec{r}) =
\frac {\nabla_i {\vec J}^{{\rm tr},i}(\vec{r}) \times \hat{m}}{m}. 
\label{deltatrnoso}
\eeq
When Eq.[~\ref{deltatrnoso}] is inserted in Eq.[~\ref{collectivedynamics2}]
it implies a contribution to the local rate of spin-density change in any small volume
proportional to the net flux of spin current into that volume; in other words it implies that
the {\em bookkeeping} theory of spin-transfer applies locally, a property that 
can be traced in this instance to the LSDA of SDFT.  This observation helps explain why
a simple spin-transfer argument \cite{Fernandez-Rossier} is able to account for the 
influence of a current on spin-waves in a homogeneous ferromagnet \cite{BJZ}.  
When spin-orbit interactions are present,
Eq.[~\ref{deltatrnoso}] is no longer valid.
 
Eqs.[~\ref{collectivedynamics2}] and [~\ref{deltatr}] provide explicit expressions for the 
effective magnetic fields that drive magnetization precession at each point in space and time.
Using these equations it is possible to explore the consequences of spatial variation in 
spin-transfer torque magnitude and direction, and of spin-orbit interactions.  These have a 
dominant importance in ferromagnetic semiconductors \cite{Koenig}.  
\begin{figure}
\vspace{-0.3cm}
\centerline{\psfig{figure=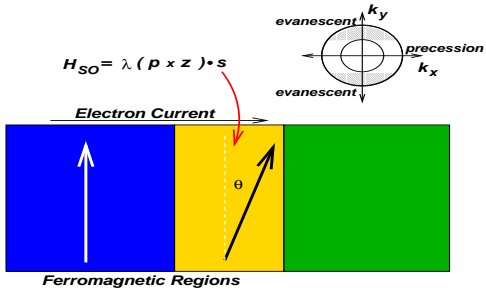,height=1.5in,width=2.5in}}
\caption{Toy model described in the text, a 2DEG with ferromagnetic 
regions.  In our calculations we apply periodic boundary conditions in the 
transverse $\hat{y}$ direction. A spin-transfer torque is present when the two magnetization directions are 
not aligned.  The inset shows the Fermi surfaces of the two ferromagnets in which are 
identical in the absence of spin-orbit coupling and and indicates the transverse 
channel $k_y$ range over which one of the two Schroedinger equation solutions is an evanescent spinor.
The Schroedinger equation solutions for electrons incident from $x \to -\infty$ can be 
solved by elementary but tedious calculations in which the spinors and their derivatives 
are required satisfy appropriate continuity conditions at the interfaces.
}
\label{fig2}
\end{figure}
\noindent 
{\em Toy-Model Calculation:} 
We illustrate our theory by evaluating
$\vec{m}^{\tr}(\vec{r})$ for a toy model containing a 
ferromagnetic two-dimensional electron system with Rashba spin-orbit interactions.
The model system, illustrated in Fig.[2], is intended to capture key features of 
the spin-transfer effect.  We take the width of the pinned magnet to infinity,
neglect the paramagnetic spacer that is required in practice to eliminate 
exchange coupling between the two magnets, and assume for 
simplicity that there is no band offset between the two ferromagnets and 
that the two exchange fields are equal
in magnitude.  Current flows from the pinned magnet, through the free magnet,
into a paramagnetic metal that functions as a load. 
The spin-orbit interaction is assumed to be confined to the free
magnet region \footnote{
To ensure Hermiticity we write $H_{SO}=
\left(\{\lambda,\vec{p}\}\times\hat{z}\right)\cdot \vec{s}$, where the symbol $\{\cdot , \cdot\}$ 
denotes the operator anticonmutator .
}
For this model we evaluated $\vec{m}^{\tr}(\vec{r})$ in
a current-carrying system using the Landauer-B\"uttiker approach \cite{Landauer}.
In the linear response regime this requires that the Schroedinger equation
be solved for electrons incident from the left at the Fermi energy 
in all transverse channels.

It is helpful at this point to make contact with the
usual description of spin-transfer.  In its simplest version, spin-transfer theory  
assumes complete transfer, {\it i.e.} that the incoming current is spin-aligned in the fixed magnet direction 
and the outgoing current is spin-aligned in the free magnet direction.
To the extent that the complete transfer assumption is valid, the torque is 
in the plane defined by the 
two magnetization orientations, which we refer to as the transfer plane.
Microscopically \cite{TransferTheory}
the component of the outgoing current perpendicular to the transfer plane 
is expected to be very small because of 
interference between precessing magnetizations in different channels.  

It follows from Eq.[~\ref{deltatr}] that the spatially averaged spin 
orientation of the transport electrons is should be 
approximately perpendicular to the transfer plane.
It can be verified that this is indeed true by 
directly evaluating $\vec{m}^{\tr}(\vec{r})$. 
This simple intuitive argument is not exact, however. In particular, the
incoming spin current is not necessarily polarized along the
pinned magnet magnetization, because of interference between
incident and reflected quasiparticle waves that complicate
the spin-transfer torques and also because it 
fails to account for electrons that are described by
spinors with evanescent components.
(See the inset of Fig.[~\ref{fig2}]).  In any microscopic calculation
these effects and others conspire to produce a relatively small component of
the torque that is perpendicular to the transfer plane, and correspondingly
to a component of $\vec{m}^{\tr}(\vec{r})$ that is in the transfer plane.

\begin{figure}
\vspace{-1.6cm}
\centerline{\epsfig{figure=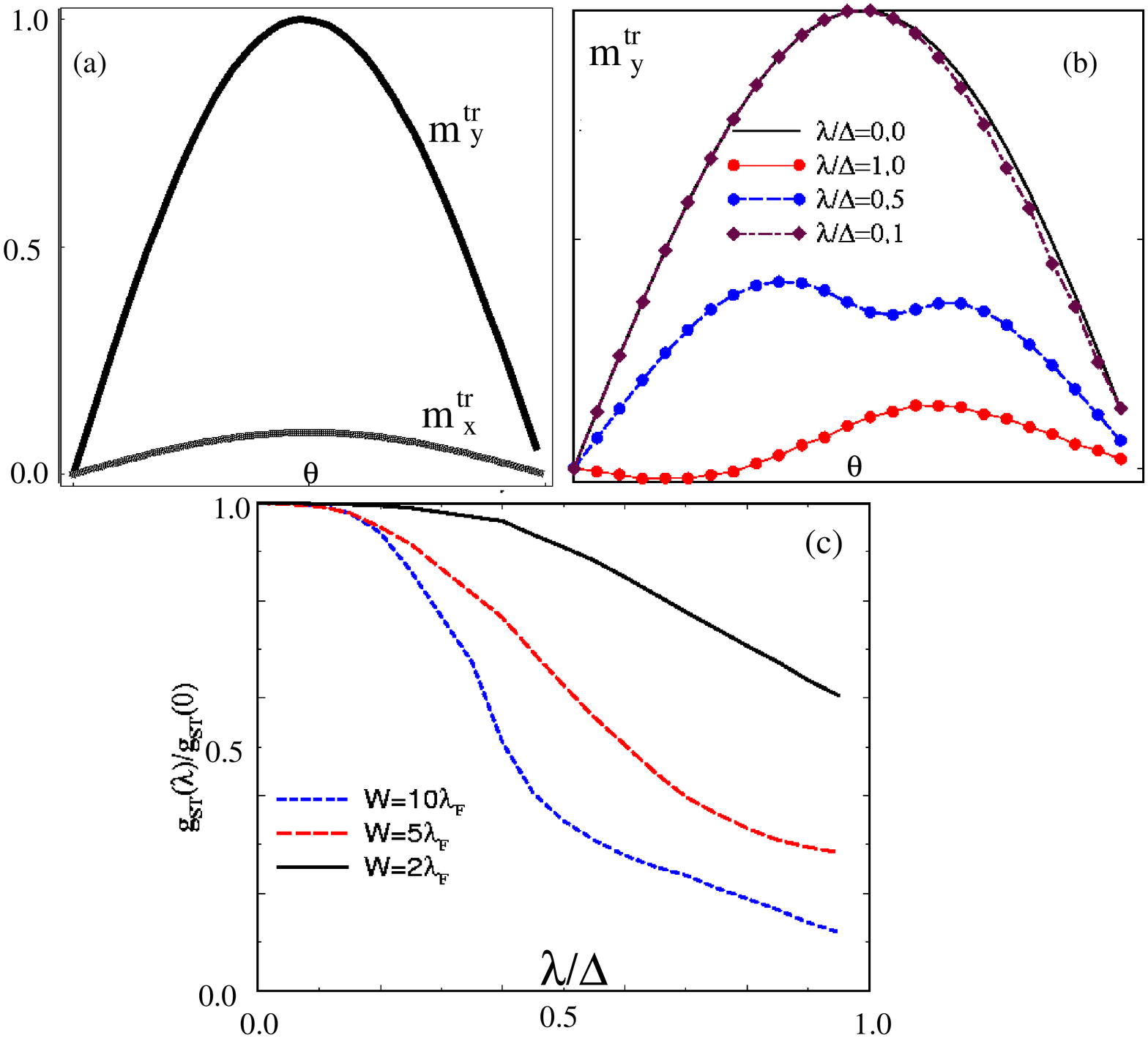,height=3.0in,width=3.3in}}
\caption{(a) Transport spin density per unit current in the case 
without spin-orbit interaction. $m^{\tr}_y$ is the component
perpendicular to the transfer plane while $m^{\tr}_x$ is the smaller 
component in the transfer plane that is
contributed by evanescent spinors. Both components are normalized to the maximum $m^{\tr}_y$
which occurs for $\theta=\pi/2$.
(b)Non-equilibrium spin density per unit current perpendicular to the transfer plane 
for different spin-orbit interaction strengths.  It follows that from these 
results that the spin-transfer torque is reduced in efficiently and altered in 
angle dependence by spin-orbit interactions. 
(c)Spin transfer efficiency, $\g_{\rm ST}$, normalized by the ST efficiency in the absence of spin-orbit coupling, 
as a function of the spin-orbit strength, for several widths of the free magnet. The spin-transfer 
effect becomes weak when the spin-orbit splitting is comparable with the 
exchange splitting.}
\label{fig3}
\end{figure}

In Fig.[~\ref{fig3}(a)] we plot values of 
$\vec{m}^{\tr}(\vec{r})$ per unit current averaged over the free magnet space  
as a function of the angle between the two magnetization orientations, 
in the case without spin-orbit interaction.  We have taken the free magnet orientation
be the $\hat{z}$ direction and the pinned magnet to be in the $\hat{z}-\hat{x}$ plane 
with polar angle $\theta$.

When spin-orbit interactions are included, the strength of the 
spin-transfer torque must be evaluated using the transport spin densities. 
The {\em bookkeeping} argument, based on total
spin conservation, is no longer valid. The quasiparticle spins not only are no longer 
conserved due to momentum-dependent effective magnetic fields that represent
spin-orbit coupling.  As we see in Fig.[~\ref{fig3}], the spin-transfer effect is not only reduced in 
magnitude but its dependence on $\theta$ no longer approximates the 
simple complete transfer expression. A measure of how the 
effect is destroyed by the spin-orbit interaction is given by the magnitude of the spin 
transfer efficiency $\g_{\rm ST}$, defined as the value of the in plane torque per unit current
at the optimum geometry, $\theta=\pi/2$. In Fig.[\ref{fig3}(c)] we show the efficiency 
as a function of the spin orbit interaction strength. We see that when the spin-orbit 
interaction strength is comparable to the exchange spin splitting the effect is 
strongly reduced except for the case of extremely thin layers.
In conclusion we have presented a formalism that
allow us to evaluate the interplay between transport currents and 
magnetization dynamics in very general circumstances. 
This formalism can address open issues in transport theory including 
the possible importance of incoherent nanomagnet magnetization dynamics,
and the influence of the spin-orbit interactions that are expected to be most
important in diluted magnetic semiconductor ferromagnets.  This work was 
supported by the Welch Foundation and by the National Science Foundation 
under grant DMR0115947.  We acknowledge helpful interactions with Gerritt Bauer,
Jack Bass, Joaquin Fernandez-Rossier, and Maxim Tsoi.


\begin{thebibliography}{99}
\bibitem{caveatonlongitudinal} The transport orbitals will also, in general, contribute
to the total spin-density component in the direction of the magnetization.  This 
effect alters the exchange field along the magnetization direction and does not 
produce a spin torque.
\end{thebibliography}
\end{document}